# Picosecond cubic and quintic nonlinearity of Lithium Niobate at 532 nm

Hongzhen Wang[1], Georges Boudebs[1] and Cid B. de Araújo[2,*]

*[1] Laboratoire de Photonique d'Angers, LPHIA, UNIV Angers, Université Bretagne Loire, 2 Boulevard Lavoisier, 49045 Angers France.*
*[2] Departamento de Física, Universidade Federal de Pernambuco, 50670-901 Recife, PE, Brazil*

*Corresponding author. E-mail: cid@df.ufpe.br

**Abstract**

The nonlinear (NL) optical response of bulk lithium niobate ($LiNbO_3$) was investigated at 532 nm using the second harmonic of a Nd:YAG laser delivering pulses of 12 ps. The experiments were performed using the D4σ method combined with the conventional Z-scan technique. Two- and three-photon absorption coefficients equal to $0.27$ cm/GW and $2.5 \times 10^{-26}$ m$^3$/W$^2$, respectively, were determined. The NL absorption processes were due to transitions from the valence to the conduction band and to free-carriers absorption. The third- and fifth-order NL refractive indices were $n_2 = (2.5 \pm 0.6) \times 10^{-19}$ m$^2$/W and $n_4 < 5.5 \times 10^{-36}$ m$^4$/W$^2$. The present results give support for previous experiments that indicate possible fifth-order processes in bulk samples and channel waveguides fabricated with $LiNbO_3$.

# I. INTRODUCTION

Metal-oxides such as Potassium Niobate (KNbO$_3$), Sodium Niobate (NaNbO$_3$), and Lithium Niobate (LiNbO$_3$) crystals present interesting properties such as piezoelectricity, electro-optic and nonlinear (NL) optical behavior. In particular, Lithium Niobate (LiNbO$_3$) has received great attention in the past four decades because of their linear and NL optical properties that are very appropriate for devices based on bulk and surface acoustic wave devices, second harmonic generation, optical waveguide modulators, among other applications [1-3]. Indeed, LiNbO$_3$ based waveguide modulators are the primary candidates for applications in optical fiber telecommunication networks and cable television.

The design and optimization of photonic and electro-optical devices motivated many groups in the past to characterize the optical properties of LiNbO$_3$ from the near-infrared to the blue range using lasers operating in various temporal regimes. The majority of the experiments were dedicated to the study of the second- and third-order optical susceptibilities, $\chi^{(2)}$ and $\chi^{(3)}$, respectively. While the works related to the second-order response of LiNbO$_3$ are in reasonable agreement, the measurements related to $\chi^{(3)}$ and high-order nonlinearities still deserve further analysis.

The third-order electronic response of LiNbO$_3$ was first investigated by DeSalvo et al. [4] using a laser operating at 532 nm (2.34 eV) with pulses of 22 ps, applying the Z-scan technique [5]. They report a third-order refractive index, $n_2 \propto Re\chi^{(3)}$, of $(440\pm70) \times 10^{-14}$ esu, and two-photon absorption (2PA) coefficient, $\beta \propto Im\chi^{(3)}$, of $(0.38\pm0.08)$ cm/GW. One year later, Li et al. [6] reported $n_2 = (5.3 \pm 0.5) \times 10^{-19}$ cm$^2$/W and $\beta = (0.25 \pm 0.6)$ cm/GW also at 532 nm. Ganeev et al. [7] working at 532 nm, with pulses of 55 ps, reported $n_2 = 3.4 \times 10^{-19}$ m$^2$/W for light propagating along the z-axis and $\beta = 0.21$ cm/GW.

More recently, Cherchi et al. [8] analyzing the efficiency of second harmonic generation in surface periodically-poled LiNbO$_3$ excited by 25 ps pulses in the near-infrared, observed reduction in the harmonic generation conversion efficiency that was attributed to three-photon absorption (3PA) processes. Working at 770 nm (1.61 eV) the authors estimate a 3PA coefficient, $\gamma \propto Im\ \chi^{(5)}$, of $4.5 \times 10^{-28}$ m$^3$/W$^2$ and negligible value of $n_4 \propto Re\ \chi^{(5)}$. Three years before, Stepanov et al. [9] investigating THz generation by optical rectification in LiNbO$_3$ excited at 800 nm (1.55 eV), also suggested that 3PA was responsible for the saturation of the THz signal being generated. However Cherchi et al. [8] and Stepanov et al. [9] could not make precise measurements of the NL absorption coefficients due to the intrinsic limitation of the techniques used.

Beyer et al. [10] also measured 2PA coefficients of LiNbO$_3$ from 388 to 800 nm, using 0.24 ps laser pulses. In the green region (at 514 nm) it was measured a 2PA coefficient of 0.1 cm/GW. High-order nonlinearities were not investigated in [10].

The main motivation for the present work was the investigation of the relative contribution of 2PA and 3PA when light at 532 nm is incident on LiNbO$_3$. The experiments were made with laser intensities varying from 10 to 150 GW/cm$^2$ using the D4σ method [11-15] combined with the Z-scan technique that allowed precise NL measurements.

This paper is organized as follows: In Section II we present some characteristics of the LiNbO$_3$ sample and a description of the experimental setup as well as the NL technique used. The NL refractive index and NL absorption measurements, at different laser intensities, are presented in Section III with an evaluation of the results. In Section IV a summary of the results and conclusions are presented.

## II. EXPERIMENTAL DETAILS

Undoped congruently grown $LiNbO_3$ crystals (z-cut) of good optical quality were obtained from Crystals Technology, Inc. The samples with 0.55 mm thickness have a composition uniformity of (48.38 ± 0.01) mol % of $Li_2O$ and present large optical transmission window from ≈315 to ≈4600 nm. For the NL measurements, we used the second harmonic of a Nd: YAG laser at 532 nm (12 ps, 10 Hz). The D4σ - Z-scan technique was applied to characterize the intensity dependent refractive behavior of the samples and the *open-aperture* Z-scan technique was used to determine the NL absorption coefficients. All measurements were performed at room temperature with linearly polarized light incident perpendicularly to the sample's surface.

The setup used for the D4σ - Z-scan measurements is illustrated in Figure 1.

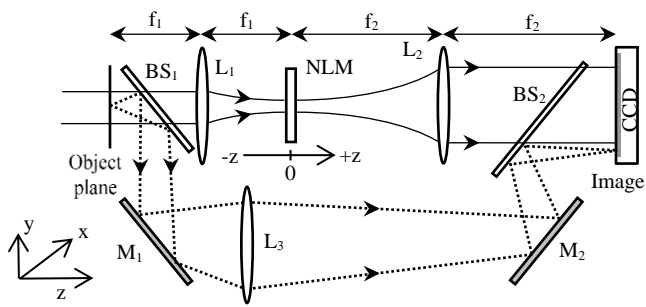

Fig.1. The 4f-imaging system. The sample is moved along the beam direction around the focal plane ($z=0$). The labels refer to: lenses ($L_1$, $L_2$ and $L_3$), beam splitters ($BS_1$ and $BS_2$), CCD camera (CCD) and mirrors ($M_1$ and $M_2$).

The sample, mounted on a translation stage, is moved along the beam direction (z axis) in the focus region. For each 1 mm step of the motor an image of the light pulses transmitted by the sample is recorded by a CCD sensor. The response of the CCD to optical power was verified to be linear. A second arm was used to monitor the fluctuations of the incident laser pulses (via lens $L_3$). The 4f-imaging system composed of two 20 cm focusing lenses ($f_1=f_2$) is described with basis on Fourier optics. Moving the sample in the focal region allows determination of the material's NL parameters because the induced NL phase-shift and/or NL absorption originate changes in the transmitted beam profile. The imaging system is aligned to obtain a magnification equal to 1.0 such that it is possible to characterize accurately the spatial profile of the beam at the entry of the experimental setup, in the linear regime (for low incident laser intensity). By using the optical transfer function for the free-space propagation over a finite distance and considering the phase transformations due to the 4f-system, we simulate the propagation of the beam from the object to the image plane taking into account the transmittance of the NL medium positioned at each motor step used for the Z-scan measurement as in [16]. The samples' transmittance is considered according to the evolution equations for the NL phase-shift and intensity in the thin sample approximation. Fits of the beam waist relative variation (BWRV) spatial profile of the output beam provide the NL refractive index of the sample by comparing the numerical and experimental data. On the other hand, the NL absorption coefficient is obtained from the fits of the conventional *open aperture* Z-scan profile. For each measurement, two sets of acquisitions are performed. The first set is in the NL regime and the second one in the linear regime, obtained by reducing the incident laser intensity; this is necessary to remove from the NL measurements the diffraction, diffusion and/or imperfection contributions due to sample inhomogeneity. Moreover, as previously mentioned, an image of the entry plane of the 4f-system allowed the characterization of the circular aperture object, with approximately 1.6 mm radius, that is considered in the computer simulation. As shown in [16] this procedure allows high accuracy measurements even for relatively high absorption and large phase-shift when the BWRV signal is no more linear with the incident laser intensity. The main source of uncertainty in the present experiments is due to the absolute measurement of the laser pulse energy because the accuracy of the joulemeter used is about 10%. To be self-consistent, in the D4σ method we use a top-hat beam in the 4f-imaging system and we acquired, on the CCD, the intensity profile I(x,y) after the interaction of the light beam with the NL material. The acquisitions obtained for each z position allow to compute the first- and the second-order moments of I(x,y), therefore, the broadening and the narrowing of the aperture radius versus z. The method gives four times the standard deviation of the spatially limited intensity distribution. The reading of the image through a CCD allows correction for point-beam instability following the centroid of the image position while calculating the first-order moment of the intensity distribution. The simulation of the whole image formation processing gives the NL parameters even in the presence of high NL absorption. No simplified formulas are used. More details on the D4σ method are given in refs. [11-16].

We remark that an advantage of using top hat beams (binary circular objects) is to be sure about the beams spatial extension while defining the object in the data acquisition program. Gaussian beams are less accurate to determine because of the inherent noise that comes with the images given by the CCD in which the wings of such beams are embedded into noise. Using circular binary objects allows to remove all the noisy pixels where the intensity is less than a treshold value of the normalized image profile because no light is transmitted outside the circular aperture. Note that if the threshold is set to a given value, the normalized object should not contain areas where the intensity is lower than this value. It is important to remark that the simplified formulas used for some kind of laser beams (such as for Gaussian beams in [11]) can not be used here due to the relatively high NL response observed in the present experiments.

**III. RESULTS AND DISCUSSION**

**A. NL coefficients measurements considering the third-order response**

The classical measurement procedure was built to consider the response of the material described by an effective cubic nonlinearity defined by $C_2$ (m/W), the third-order NL absorption coefficient, and $n_{2eff}$ (m²/W), the effective NL refractive index. In these conditions, the amplitude transmittance of the sample is described by $T(z,u,v) = \left[1+q(z,u,v)\right]^{-1/2} \exp\left[j\Delta\varphi_{NL}^{eff}(z,u,v)\right]$, where z represents the position of the sample in the focus region. $u=x/\lambda f_1$ and $v=y/\lambda f_1$ are the normalized spatial frequencies with $\lambda$ as the incident laser wavelength. $q(z,u,v) = C_2 L I(z,u,v)$ where L denotes the thickness of the sample and $I(z,u,v)$ is the intensity of the laser beam inside the sample. The NL phase-shift is determined by $\Delta\varphi_{NL}^{eff}(z,u,v) = 2\pi n_{2eff} L I_{eff}(z,u,v)/\lambda$ where the effective intensity seen by the sample is represented by $I_{eff}(z,u,v) = I(z,u,v)\log\left[1+q(z,u,v)\right]/q(z,u,v)$. The peak on-axis NL phase-shift at the focus is $\Delta\varphi_0 = \Delta\varphi_{NL}^{eff}(0,0,0)$ which can be deduced from the BWRV signals (for more details see [11]). To obtain $n_{2eff}$, it is necessary to take into account the actual NL absorption measured at this intensity level considering $C_2$.

The results obtained by the fittings are represented by the solid lines (blue color) in Figure 2 for the NL refraction and Figure 3 for the NL absorption for three different intensities chosen in the measuring interval.

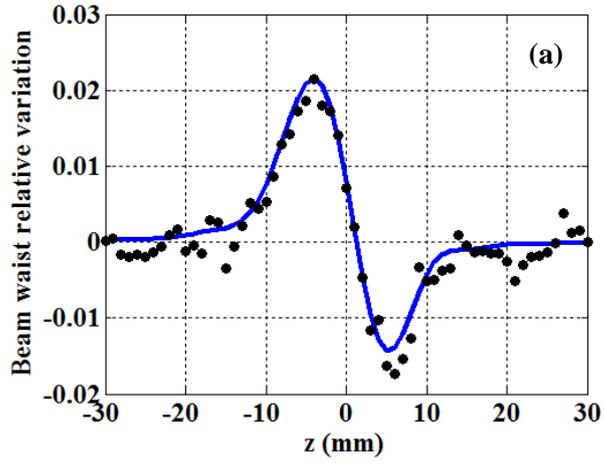

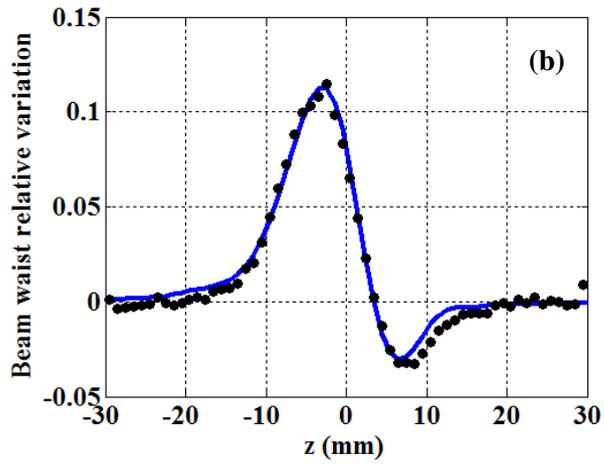

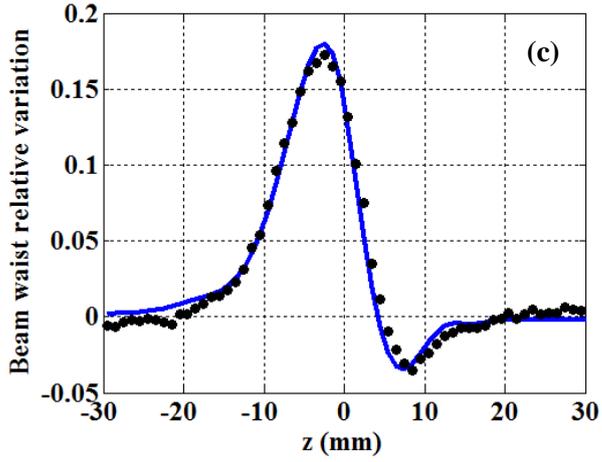

Fig. 2. (Color online) The NL refraction signal associated to the beam waist relative variation versus z for a 0.55 mm LiNbO$_3$ crystal (z-cut) measured with (a) $I_0 = 1.1 \times 10^{14}$ W/m$^2$, (b) $I_0 = 6.9 \times 10^{14}$ W/m$^2$, (c) $I_0 = 13.4 \times 10^{14}$ W/m$^2$. Laser wavelength: $\lambda = 532$ nm. The experimental data are represented by the black filled circles; the blue solid lines are the best-fitting results.

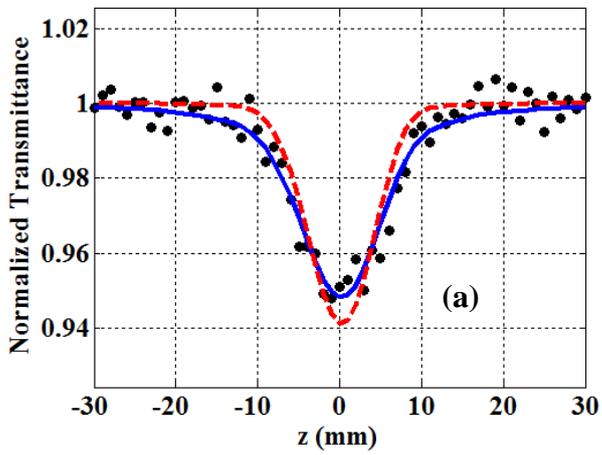

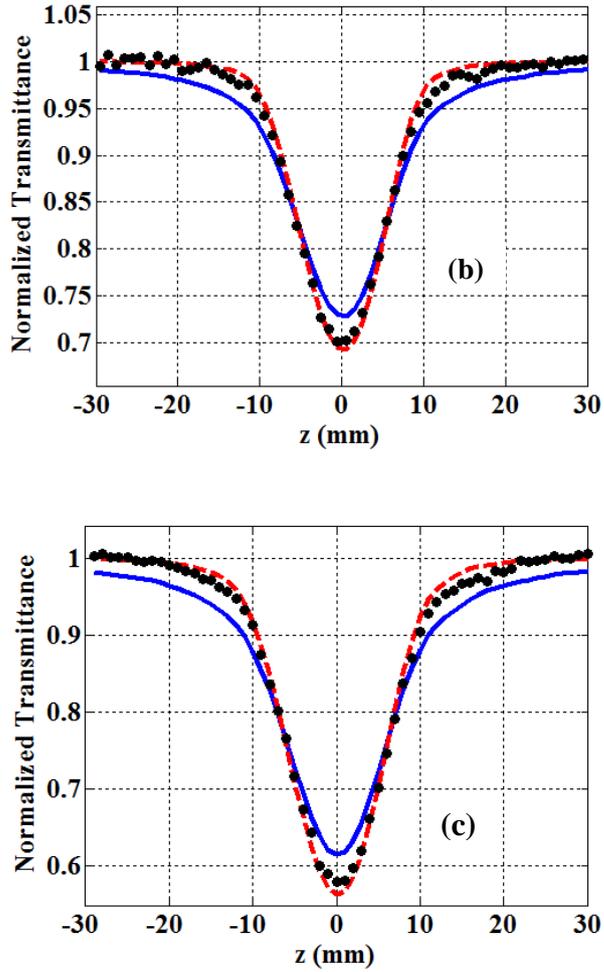

Fig.3 (Color online) The NL absorption signal versus z obtained with the same experimental parameters as in Figure 2. Experimental data: black filled circles. The blue solid lines are fits for 2PA and the dashed red lines are for 3PA.

A very good agreement between the experimental data and the fittings can be observed for the NL refraction signals (BWRV) but the NL absorption *open aperture* Z-scan profiles show a significant difference with the 2PA fitting profiles (solid lines in Figure 3). The results are summarized in Figure 4.

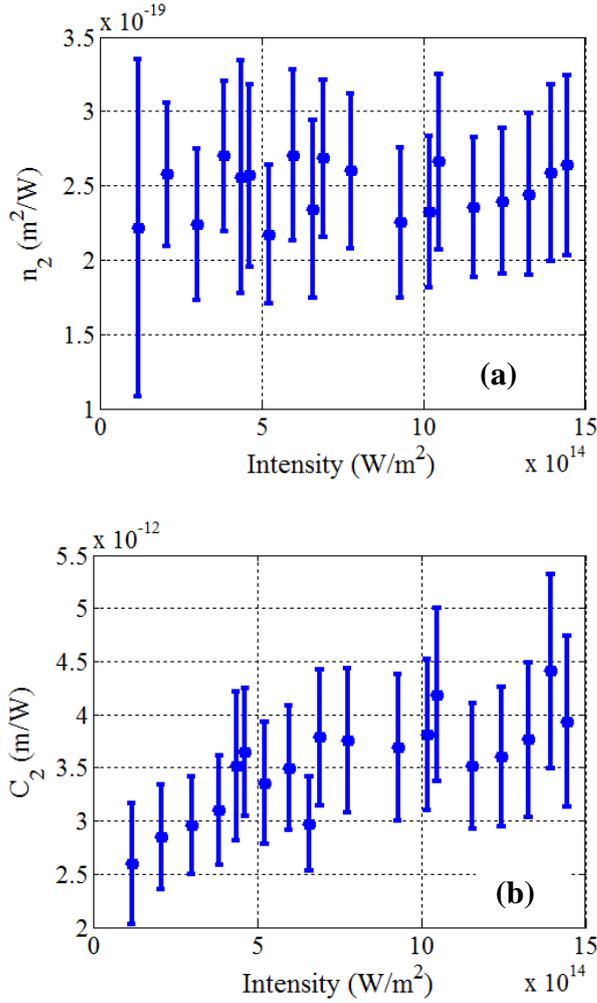

Fig. 4 (Color online) Evolution of the measured effective third-order coefficients as a function of the laser intensity: (a) the NL refraction coefficient and (b) the 2PA coefficient.

Figure 4(a) shows the behavior of $n_{2eff}$ for the various intensities showing a constant value whose average is $(2.5 \pm 0.6) \times 10^{-19}$ m²/W. Taking into account the uncertainties of the measurements, our value is in agreement with those found in the literature by the authors of [6, 7]. If one considers the simple linear relation that account for a fifth-order NL refraction coefficient: $n_{2eff} = n_4 I_0 + n_2$, the slope ($n_4 \propto \text{Re } \chi^{(5)}$) of the linear fitting is so low that it would indicate a negligible value for $n_4$ (a rough estimation according to the data of this figure and taking into account the error bars gives $n_4 < 5.5 \times 10^{-36}$ m$^4$/W$^2$).

Figure 4(b) summarizes the behavior of $C_2$ showing a noticeable increase versus the incident intensity. The average value of $C_2$ calculated up to intensities of $1.5 \times 10^{15}$ W/m$^2$ is $(3.5 \pm 0.7) \times 10^{-12}$ m/W in good agreement with [6, 7]. A comparison with the result of [10] indicates a larger 2PA coefficient in the present experiment. The discrepancy may be due to the difference between the pulses duration in both experiments. Typically the NL coefficients measured in the picosecond regime are higher than those obtained with femtosecond pulses that normally excite only the pure electronic transitions.

The intensity dependent value of $C_2$ and the imperfect agreement of the 2PA fittings shown in Figure 3 motivated us to investigate higher order NL absorption as will be presented in the next subsection.

## B. NL absorption measurements considering higher-order response

We assume that, under the slowly varying envelope and thin sample approximations, the evolution of the optical intensity, I, as a function of $z'$, the light propagation axis inside the material, obeys the equation:

$$\frac{dI}{dz'} = -C_N I^N \qquad (1)$$

where N is an integer to be determined. Equation (1) is identical to the evolution equation for the intensity when the losses are due to N-photon absorption (NPA) with $C_N$ being the corresponding NL absorption coefficient. Considering the boundary conditions $I(z' = 0) = I_0(z, u, v, t)$ and $I(z' = L) = I_L(z, u, v, t)$, the solution of Equation (1) is:

$$I_L(z, u, v, t) = \frac{I_0(z,u,v,t)}{\{1+(N-1)C_N L[I_0(z,u,v,t)]^{N-1}\}^{1/(N-1)}} \quad (2)$$

Of course for N=2, Eq (2) reduces to the squared modulus of the amplitude transmittance T defined in the subsection A.

For the *open-aperture* Z-scan profile, Equation (2) is integrated numerically over space and time (over u, v and t) assuming a Gaussian temporal profile and the actual spatial profile of the beam at the entry (after propagation to the z position in the focal region). The numerical integration at this level for each z position using Equation (2) is advantageous because it avoids approximations that assume relatively low NL absorption correction in the far-field, i.e.: $|(N-1)C_N L[I_0(z, u, v, t)]^{N-1}| \leq 1$. Thus, performing the *open-aperture* Z-scan experiment, the value of $C_N$ can be deduced, for any regime by fitting the normalized transmittance data profile.

## C. Co - existence of 2PA and 3PA processes

$LiNbO_3$ exhibits a wide transparency window in the visible but there is a strong absorption band from ≈4.2 to ≈7.3 eV [17, 18]. Therefore, two- and three-photon absorption processes are expected by considering the large intensities of the incident laser with wavelength 532 nm (2.34 eV). For this reason the data in Figure 3 at low-intensity was well described by 2PA (the blue solid line) considering $C_2 = 0.27$ cm/GW, while at moderate and large intensities the data is better adjusted by considering 3PA (the red dashed line) as the dominant process with $C_3 = 2.5 \times 10^{-26}$ m$^3$/W$^2$.

Figure 5 shows the intensity dependence of $C_3$ which results in a constant value particularly for intensities larger than $5 \times 10^{14}$ W/m$^2$.

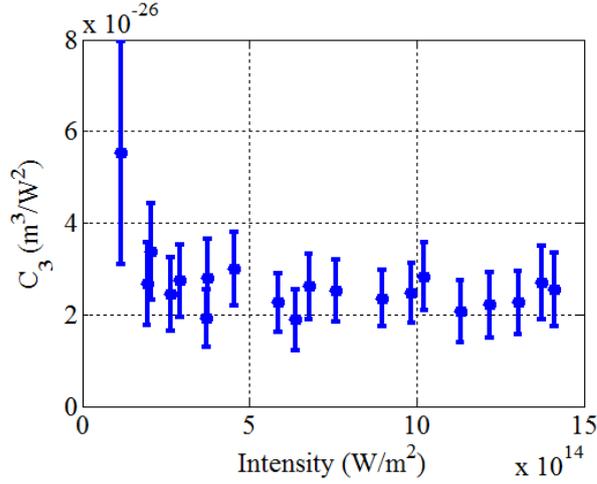

Fig. 5. (Color online) Measured 3PA coefficient, $C_3$, as a function of the incident intensity. The measured values were obtained by the fitting of the same experimental *open aperture* Z-scan profiles used in Figure 4(b) but performed using Equation 2 with N=3. Note the constant value obtained on the large interval going from 10 to 140 GW/cm$^2$.

The relative contribution of the 2PA and 3PA processes can be evaluated in the following way. First, we recall that the best-fit line is the line that minimizes the squared deviations from theory given by $E = \sum_i (T_{ei} - T_{pi})^2$, where $T_{pi}$ is the predicted value on the line for a given number position i, and $T_{ei}$ is the actual value measured for that given i. Then, we defined the Mean Vertical Deviation as $MVD = \frac{E}{N}$ with N being the total number of acquisitions (generally 61) for a given intensity corresponding to the Z-scan profile. The intensity dependence of MVD is illustrated by Figure 6. Notice that the MVD globally increases with the laser intensity when considering the 2PA process while the fitting is much better considering 3PA as the main contribution for high and moderate intensities (see the dashed lines in Figure 3).

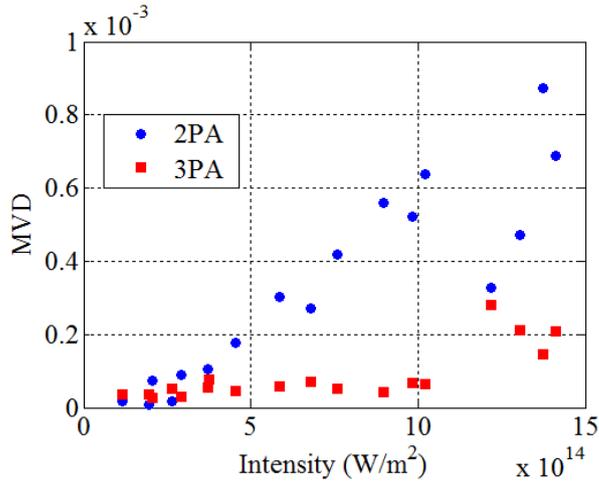

Fig. 6 (Color online) The mean value of the vertical deviations between the data points and the fitting as a function of the laser intensity. 2PA (blue filled circles) and 3PA (red filled squares).

Therefore, the results shown in the Figure 3 and Figure 4 indicate that the two NL absorption processes, 2PA and 3PA, occur for intensities lower than $5 \times 10^{14}$ W/m$^2$ but the 2PA process is dominant. The 3PA process is dominant for laser intensities larger than $5 \times 10^{14}$ W/m$^2$ because the electron population in the conduction band becomes large and therefore the one-photon free-carriers absorption becomes relevant. That may explain the different conclusions by the authors of [6, 7] and [8, 9].

## IV. SUMMARY

In the present paper we demonstrate that congruent LiNbO$_3$ in the presence of strong picosecond lasers operating at 532 nm may present two-photon absorption from the valence to the first conduction band and three-photon absorption. The three-photon absorption process is a two-step excitation process (two-photon absorption without real intermediate state to the first conduction band followed by one-photon absorption by free-carriers). The 2PA coefficient $C_2$ = 0.27 cm/GW was determined at relatively low-intensity while at moderate and large intensities the dominant process was found to be 3PA with $C_3 = 2.5 \times 10^{-26}$ m$^3$/W$^2$. The third- and fifth-order NL refractive index were respectively $n_2 = (2.5 \pm 0.6) \times 10^{-19}$ m$^2$/W and $n_4 < 5.5 \times 10^{-36}$ m$^4$/W$^2$.

**ACKNOWLEDGEMENTS**

This work was performed in the framework of the National Institute of Photonics (INCT de Fotônica) project. We acknowledge financial support from the Brazilian agencies Conselho Nacional de Desenvolvimento Cientifico e Tecnológico (CNPq) and the Fundação de Amparo à Ciência e Tecnologia do Estado de Pernambuco (FACEPE). Cid B. de Araújo appreciates the hospitality of the Université d'Angers where the experiments were made and the financial support from the LUMOMAT and the NNN-TELECOM programs.